\documentclass[amsmath,superscriptaddress,showpacs,prl,twocolumn] {revtex4-2}
\usepackage{bm}
\usepackage{bbm}
\usepackage{enumerate}
\usepackage{graphicx}
\usepackage[dvips]{epsfig}
\usepackage{epsf}
\usepackage{physics}
\usepackage{xcolor}
\usepackage{bbold}
\usepackage[normalem]{ulem}
\usepackage[caption=false]{subfig}
\usepackage{url}

\makeatletter
\newcommand{\Rmnum}[1]{\expandafter\@slowromancap\romannumeral #1@}
\makeatletter

\allowdisplaybreaks

\begin{document}
\title{Control of spin waves by spatially modulated strain}
\author{Ankang Liu}
\affiliation{Department of Physics and Astronomy, Texas A\&M University, College Station, Texas 77843-4242, USA}
\author{Alexander M. Finkel'stein}
\affiliation{Department of Physics and Astronomy, Texas A\&M University, College Station, Texas 77843-4242, USA}
\affiliation{Department of Condensed Matter Physics, The Weizmann Institute of Science, Rehovot 76100, Israel}

\begin{abstract}

We suggest using spatially modulated strain for control of a spin wave propagating inside a bulk magnet. The modulation with the wave vector $\bm q=2\bm k$, by virtue of magnetoelasticity, mixes spin waves with wave vectors near $\bm k$ and $-\bm k$. This leads to lifting the degeneracy of the symmetric and antisymmetric eigenstate combinations of these waves. The resulting picture reminds one of a tunneling particle in a symmetric double-well potential. Here, a moving spin wave being subjected to the $2\bm k$-lattice modulation after some time alters its propagation direction to the opposite one, and so on. The effect can be utilized for the control of the spin-wave propagation that can be useful for spintronic and magnonic applications. The control may include a delay line element, filtering, and waveguide of the spin waves.

\end{abstract}


\maketitle

\emph{Introduction.}--In this letter, we suggest using a spatially modulated strain of the lattice for the control of spin waves (magnons) propagating inside bulk magnets through the magnetoelasticity. Actually, interaction between spin waves and mechanical excitations via magnetoelasticity has been studied for a long time, starting from the works by Kittel \cite{kittel1958interaction}. Both, the generation of the hypersonic waves by excitation of the spin system in ferromagnets \cite{bommel1959excitation} and, conversely, the generation of spin waves by pumping microwave phonons \cite{pomerantz1961excitation,matthews1964phonon} were discussed. Different aspects of the magnetoacoustic resonance and parametric excitation of magnetostatic and elastic modes have been considered \cite{spencer1958magnetoacoustic,comstock1963aparametric,comstock1963bparametric,joseph1970dependence}.

In recent years, parametric pumping of spin waves by acoustic waves has been experimentally realized \cite{chowdhury2017nondegenerate} as well as elastically driven ferromagnetic resonance \cite{weiler2011elastically, dreher2012surface}. An enormous increase in the amplitude of the magnetization precession in a ferromagnetic layer embedded into a phononic resonator was observed in Ref. \cite{jager2015resonant} when the frequencies of magnetization precession and phonons were equal. Next, traveling acoustic waves on the surface of a piezoelectric crystal resonantly excite traveling surface spin waves in an adjacent thin-film ferromagnet. These measurements provide a spectroscopy technique for the surface spin waves \cite{gowtham2015traveling}. Recently, a nonreciprocal surface acoustic wave propagation due to the magneto-rotation coupling was also demonstrated experimentally \cite{xu2020nonreciprocal}.

Here, unlike most of the works cited above, we exploit not the dynamics of phonons but rather the \emph{spatial} modulation of the lattice. The deformation of the lattice modulates the spin exchange between magnetic atoms, which in turn acts as a scattering potential for the spin waves. The intensity of the spin wave is assumed to be weak, so that magnons are described by linearized equations, and no interconversion between phonons and magnons \cite{li2021advances} will be considered. The main idea looks as follows: The spin wave is a degenerate excitation, i.e., energies of the symmetric and antisymmetric eigenmodes with wave vectors $\pm k$ are degenerate. A spatial modulation caused by either a static strain \cite{chen2021strain} or a standing acoustic wave \cite{chowdhury2017nondegenerate} with $q=2k$ lifts this degeneracy. In the presence of a static $2k$ modulation, the picture reminds one of a particle in a symmetric double-well potential. Tunneling, as is well known, lifts the degeneracy of the energy levels in the double well. If originally a particle is located in one of the wells, as a result of tunneling it starts to oscillate between the two wells with a frequency proportional to the level splitting. Here, the strain leads to a similar effect. Suppose, initially, there is a free right-moving spin wave in the magnetic system and, then, at a certain moment, a strain modulation is switched on. (For example, a spin-wave packet runs inside the magnet when a strain is switched on.) As we have shown here, the originally right-moving spin wave being subjected to the deformation, after some time, will alter its motion to the left-moving propagation, and so on. Thus, a direct propagation of the spin wave changes into a \emph{to-and-fro} motion.

\emph{Equations of motion.}--Here, for the purpose of simplicity, we consider a layered antiferromagnet with the spin wave propagating in the direction across the layers. This simple geometry allows to illustrate the main idea. Note, however, that the method of controlling the propagation of the spin waves proposed in this letter is general and applicable to any magnetic system. In van der Waals layered systems, like CoTiO$_3$, spins in each of the layers ($xy$ planes) are arranged ferromagnetically on a graphene-like honeycomb lattice. These ferromagnetic layers are ABC-stacked along the third direction ($z$ axis). The exchange coupling between the layers is antiferromagnetic. Measurements of the magnon spectrum \cite{yuan2020dirac} allowed the Hamiltonian that best describes this system of ABC-stacked honeycomb lattices of spins to be constructed:
\begin{align}\label{H_spins_MT}
H=&\sum_{i,\delta_1}J_{\parallel}(S_{i}^{x}S_{i+\delta_1}^{x}+S_{i}^{y}S_{i+\delta_1}^{y})
\nonumber
\\
&+\sum_{i,\delta_2}J_{\perp}(S_{i}^{x}S_{i+\delta_2}^{x}+S_{i}^{y}S_{i+\delta_2}^{y}+S_{i}^{z}S_{i+\delta_2}^{z}).
\end{align}
Notations here are the same as in Ref. [\onlinecite{yuan2020dirac}]; $J_{\parallel}<0$ and $J_{\perp}>0$ are the intralayer and interlayer exchange coupling constants. The index $i$ runs over all sites of spin, while $\delta_1$ and $\delta_2$ run over the nearest neighbors within and the next-nearest neighbors between the layers, respectively. As a result, CoTiO$_3$ is actually an intralayer XY ferromagnet and interlayer antiferromagnet.

Our goal now will be to switch to continuous variables \cite{mattis2012theory,auerbach2012interacting}, instead of using the spin operators located on the lattice sites. With this in mind, we introduce $\bm{S}(\bm{r},t)$ and $\bar{\bm{S}}(\bm{r},t)$ to be spin operators for the alternating $\pm x$-ordered magnetic layers. Eventually, spin dynamics of the layered antiferromagnetic material CoTiO$_3$ will be analyzed in terms of the macroscopic quantities, namely, the total magnetization $\bm{m}\equiv\bm{S}+\bar{\bm{S}}$ and the N\'eel vector $\bm{l}\equiv\bm{S}-\bar{\bm{S}}$ \cite{auerbach2012interacting}. The derivation of the equations of motion for $\bm{m}$ and $\bm{l}$, with and without lattice deformations, is the standard one.

We take the standard parametrizations $\bm{l}=2\tilde{S}\allowbreak\times(\cos\theta\cos\phi,\cos\theta\sin\phi,\sin\theta)$ and $\bm{m}=(-m_{\theta}\sin\theta\cos\phi\allowbreak-m_{\phi}\sin\phi,-m_{\theta}\sin\theta\sin\phi+m_{\phi}\cos\phi,m_{\theta}\cos\theta)$, where $\theta^{(0)}=0$, $m_{\theta}^{(0)}=0$, and $m_{\phi}^{(0)}=0$ are the equilibrium values for this system. In CoTiO$_3$, the effective spin $\tilde{S}=1/2$. Note, in this connection, that the layered structure of this antiferromagnet improves the situation with the accuracy of the analysis of the spin dynamics at small $\tilde{S}$. The point is that the value of the intralayer ferromagnetic coupling constant $J_{\parallel}$ is much larger than that of the antiferromagnetic one $J_{\perp}$; namely, $J_{\parallel}=-4.41$ meV, while $J_{\perp}=0.57$ meV. (This choice of the coupling constants matches quantitatively well with the experimental data for the magnon spectrum \cite{yuan2020dirac}.) Because of this inequality, spins act effectively as large-spin clusters. This makes possible the description in terms of macroscopic classical variables.

After \emph{linearization,} the equations of motion for $\bm{l}$ and $\bm{m}$ decouple into two pairs $(m_{\theta},\phi)$ and $(m_{\phi},\theta)$:
\begin{align}\label{EOM_m_theta_MT}
&\dot{m}_{\theta}\approx(4\tilde{S}^{2})(-\frac{3}{8}J_{\parallel}\nabla_{-}^{2}\phi+\frac{9}{8}J_{\perp}\nabla_{+}^{2}\phi),
\nonumber
\\
&\dot{\phi}\approx(-\frac{3}{2}J_{\parallel}+9J_{\perp})m_{\theta};
\end{align}
And
\begin{align}\label{EOM_m_phi_MT}
&\dot{m}_{\phi}\approx(4\tilde{S}^{2})(-\frac{3}{2}J_{\parallel}\theta-\frac{9}{8}J_{\perp}\nabla_{+}^{2}\theta),
\nonumber
\\
&\dot{\theta}\approx(-9J_{\perp}-\frac{3}{8}J_{\parallel}\nabla_{-}^{2}-\frac{9}{8}J_{\perp}\nabla_{+}^{2})m_{\phi}.
\end{align}
Here, we have introduced a short notation, $\nabla_{\pm}^{2}\equiv\nabla^{2}\pm\partial^{2}/\partial z^{2}$; all lengths in our discussion are measured in the units of either intralayer or interlayer lattice constants and therefore are \emph{dimensionless}. In fact the form of the equations does not depend much on the microscopic details of the Hamiltonian (\ref{H_spins_MT}) and is determined by the symmetry of the system. Only the numerical prefactors are specific to the model. Note that $\theta$, $m_{\theta}$, and $m_{\phi}$ are deviations from the equilibrium values, while angle $\phi$ can be arbitrary, because the discussed system has rotational symmetry with respect to the $z$ direction.

The equations for the pair $(m_{\theta},\phi)$ lead to the acousticlike branch of the spin waves (corresponds to the Goldstone mode of the system \cite{auerbach2012interacting,takei2014superfluid}), while the equations for the pair $(m_{\phi},\theta)$ give the spectrum of the opticlike branch. Here, we are only interested in the dynamics of the acousticlike magnons.

\emph{Dynamics of $(m_{\theta},\phi)$ at static strain modulation.}--We first explain the idea behind the calculation. The deformation induced by the strain modulates the spin exchange between magnetic atoms, which in turn acts as a scattering potential for spin waves. Now, let us consider a one-dimensional problem by assuming that there is a \emph{static} deformation with only one nonzero strain tensor component $\epsilon_{zz}=\epsilon_0\cos(qz)$ (a possible realization of the static strain modulation was proposed in Ref. \cite{chen2021strain}). Here, $\epsilon_0$ is the magnitude of the strain tensor, and the wave vector $q$ describes its spatial modulation along the $z$ direction. Consequently, Eqs. (\ref{EOM_m_theta_MT}) have to be modulated to include magnetoelasticity
\begin{align}\label{EOM_1_MT}
&\dot{m}_{\theta}=J\frac{d^{2}}{dz^{2}}\phi,
\nonumber
\\
&\dot{\phi}=[G_1+G_2\cos(qz)]m_{\theta},
\end{align}
where $G_1=-3J_{\parallel}/2+9J_{\perp}$. In the discussed geometry, with the strain applied along the direction perpendicular to the layers, there is a very clear separation of the roles of $J_{\parallel}$ and $J_{\perp}$. Namely, in the above pair of equations, $J=9\tilde{S}^{2}J_{\perp}$ and $G_2=9g_2\epsilon_0$. Here, the magnetoelastic coefficient $g_2\equiv(1/c)\partial J_{\perp}/\partial c$ describes the sensitivity of $J_{\perp}$ to a modulation of the dimensionless interlayer distance $c$.

We proceed with Eqs. (\ref{EOM_1_MT}) by taking another time derivative in each of them:
\begin{align}\label{EOM_2_MT}
&\ddot{m}_{\theta}=(D_{m}+D_{m}^{(2)})m_{\theta},
\nonumber
\\
&\ddot{\phi}=(D_{\phi}+D_{\phi}^{(2)})\phi.
\end{align}
Here, we defined the operators $D_{m/\phi}\equiv JG_1d^{2}/dz^{2}$ and $D_{m}^{(2)}\equiv JG_2\cos(qz)d^{2}/dz^{2}-2JG_2q\sin(qz)d/dz-JG_2q^{2}\cos(qz)$, $D_{\phi}^{(2)}\equiv JG_2\cos(qz)d^{2}/dz^{2}$. Note that $m_{\theta}$ and $\phi$ are not decoupled, because they are connected through the relation $\dot{m}_{\theta}=Jd^{2}\phi/dz^{2}$.

To find eigenstate solutions for $\phi$, we assume that $\phi$ has a form $\phi(z,t)=e^{\pm i\omega t}\varphi(z)$, substitute this ansatz into the second line of Eq. (\ref{EOM_2_MT}), and finally, obtain a time-independent equation for $\varphi$:
\begin{align}\label{eigen_eq_varphi_MT}
-\omega^{2}\varphi=(D_{\phi}+D_{\phi}^{(2)})\varphi.
\end{align}
To solve this eigenvalue problem, we write $\varphi(z)=\sum_{k\geq0}[\mathcal{S}_k\sin(kz)+\mathcal{C}_k\cos(kz)]$. In order to find the expansion coefficients $\mathcal{S}_{k}$ and $\mathcal{C}_{k}$, we calculate matrix elements of the operators $D_{\phi}$ and $D_{\phi}^{(2)}$. We observe that there is no mixture between the basis functions $\sin(kz)$ and $\cos(kz)$; see Sec. S2 in Supplemental Material (SM) \cite{SM}.\nocite{nicklow1972lattice,wakabayashi1975lattice,weisstein2020mathieu}

We are interested in the special region of wave vectors $k\approx q/2$. In this case, we can treat the $z$-coordinate dependencies in the spirit of the parametric resonance theory. We therefore will neglect the higher harmonics, such as $\mathcal{S}_{3q/2}$ and $\mathcal{C}_{3q/2}$ (a comprehensive discussion of this important point is presented in Sec. S4 of SM \cite{SM}). Eventually, instead of a chain of coupled equations, we get a finite system of equations. Moreover, exactly at the \emph{``resonance"} defined by the condition for the wave vectors (rather than frequencies) $k=q/2$, it reduces to a pair of decoupled equations:
\begin{align}\label{eq_resonance_S_MT}
\omega^{2}\mathcal{S}_{\frac{q}{2}}=JG_1[1-\frac{1}{2}(\frac{G_2}{G_1})](\frac{q}{2})^{2}\mathcal{S}_{\frac{q}{2}}
\end{align}
and
\begin{align}\label{eq_resonance_C_MT}
\omega^{2}\mathcal{C}_{\frac{q}{2}}=JG_1[1+\frac{1}{2}(\frac{G_2}{G_1})](\frac{q}{2})^{2}\mathcal{C}_{\frac{q}{2}}.
\end{align}
The above equations lead to the split frequencies $\omega_{\mathcal{S},q/2}^{2}=JG_1[1-(G_2/2G_1)](q/2)^{2}$ and $\omega_{\mathcal{C},q/2}^{2}=JG_1[1+(G_2/2G_1)](q/2)^{2}$ for the modes $\sin(qz/2)$ and $\cos(qz/2)$, respectively.

\emph{To-and-fro motion at the resonance.}--Let us assume that initially there is a freely propagating spin wave with $\phi(z,t)=\phi_0\sin(qz/2-\Omega t+\varphi_1)$, where in the absence of the strain $\Omega=\sqrt{JG_1}(q/2)\equiv v_s(q/2)$. Next, at a moment $t=0$, the strain modulation with the wave vector $q$ switches on. In this sense, $\varphi_1$ is defined as the phase differences of the freely propagating spin wave and the strain modulation at the moment of switching on the deformation. The modulation splits the energy of the initially degenerate states. We demonstrate now that the difference between $\omega_{\mathcal{C},q/2}$ and $\omega_{\mathcal{S},q/2}$ leads to a \emph{to-and-fro} motion for the $q/2$-spin wave, exactly like in the case of a particle in the double-well with the energy levels split by tunneling. In the following part of the letter we discuss the to-and-fro motion in detail.

\begin{figure}[ht] \centerline{\includegraphics[clip, width=1 \columnwidth]{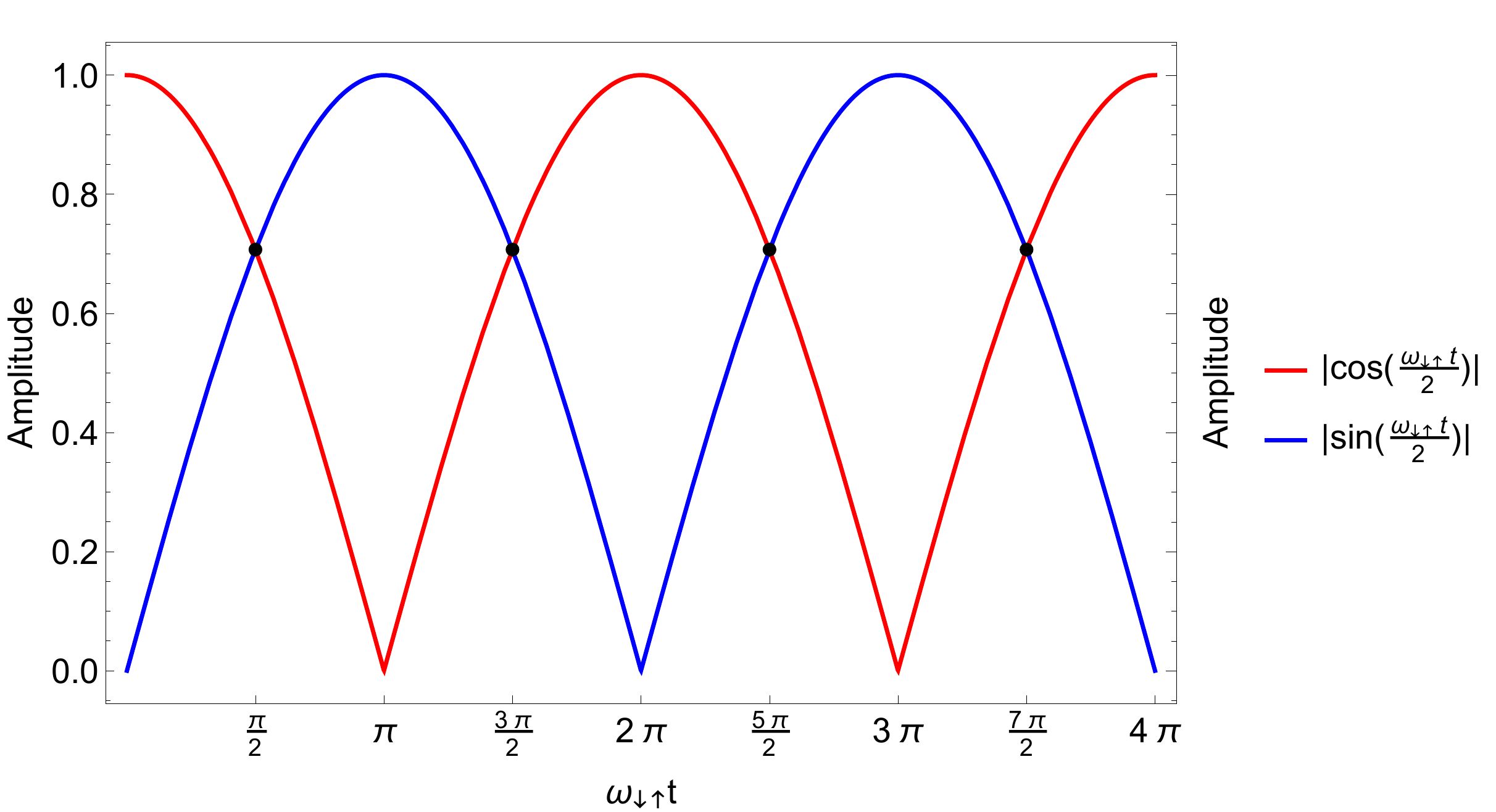}}
	
	\protect\caption{The time dependence of the coefficients of the right- and left-propagating wave components according to Eq. (\ref{phi_static_MT}).}
	
	\label{fig:static_time_dependent_coefficients}
\end{figure}

\begin{figure}[h] \centerline{\includegraphics[clip, width=1 \columnwidth]{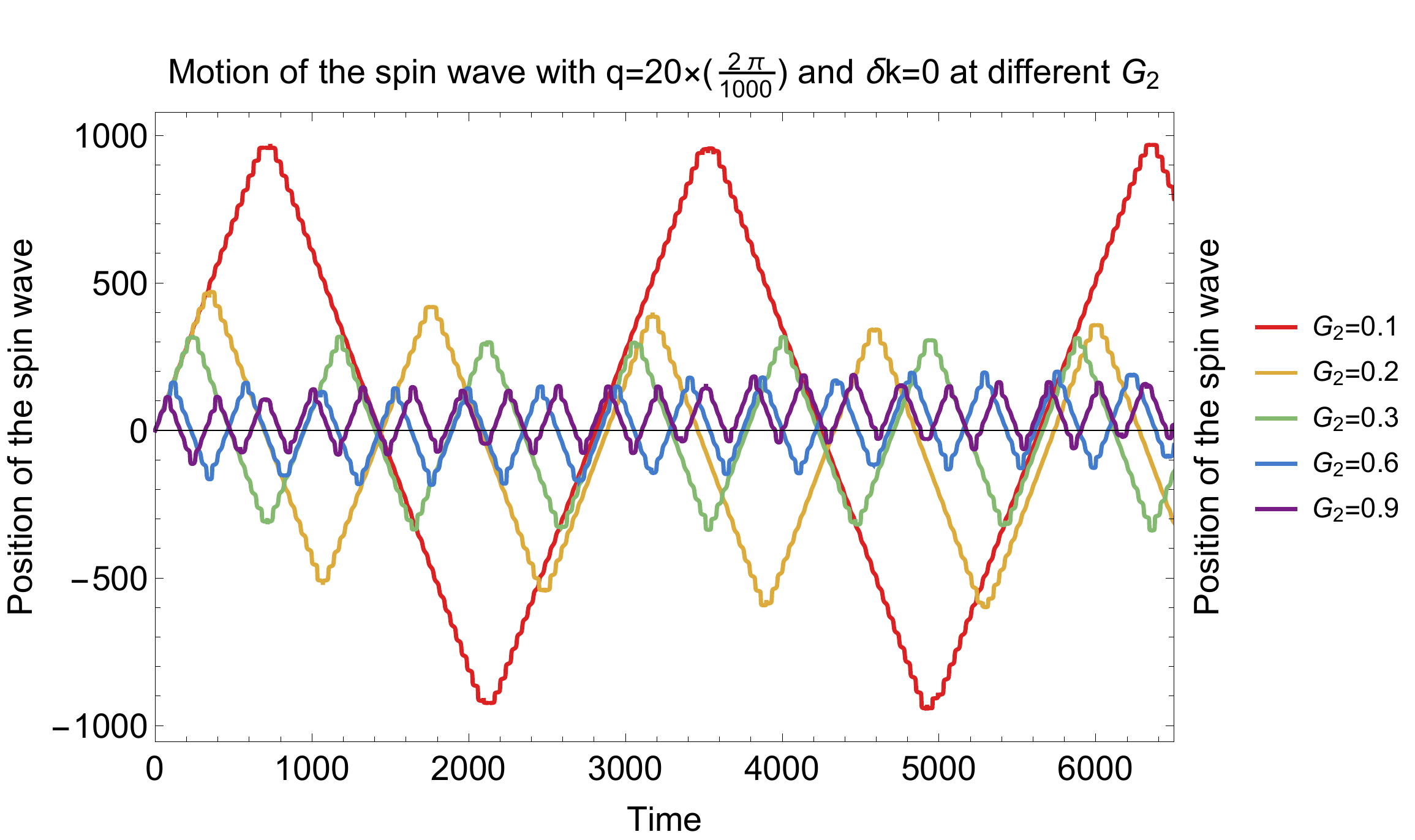}}
	
	\protect\caption{Position of spin waves at the spatial resonance condition, $k=q/2$, as a function of time for different $G_2$. Wave vector $q=20\times(2\pi/1000)$; other parameters are $J=1$, $G_1=2$, and $\varphi_1=0$.}
	
	\label{fig:static_G_2}
\end{figure}

By matching the initial conditions $\phi(z,t=0)=\phi_0\times\allowbreak\sin(qz/2+\varphi_1)$ and $m_{\theta}(z,t=0)=-J(q/2)^2\phi_0/\Omega\times\allowbreak\cos(qz/2+\varphi_1)$, and after neglecting all the small terms, one obtains
\begin{align}\label{phi_static_MT}
	\phi\approx&\phi_0\Big\{(+1)\cos(\frac{\omega_{\downarrow\uparrow} t}{2})\sin[(\frac{q}{2})z-\Omega t+\varphi_1]
	\nonumber
	\\
	&+(-1)\sin(\frac{\omega_{\downarrow\uparrow} t}{2})\cos[(\frac{q}{2})z+\Omega t-\varphi_1]\Big\}.
\end{align}
The resulting combination describes alternation between the two components propagating in the opposite directions. It works as follows: When $|\cos(\omega_{\downarrow\uparrow}t/2)|>|\sin(\omega_{\downarrow\uparrow}t/2)|$, see Fig \ref{fig:static_time_dependent_coefficients}, the right-propagating component dominates, and thus the superposition of the right- and left-propagating waves is moving toward the right, and vice versa.
For an \emph{observer} focused on a certain point of the wave it will look like \emph{to-and-fro} motion of the spin wave. (In the case of the wave packet of spin waves centered around the wave vector $k=q/2$, the packet will exhibit an alternating motion in the opposite directions, see Sec. S6 in SM [21] for the result of simulations.) As follows from Eq. (\ref{phi_static_MT}) and Fig. \ref{fig:static_time_dependent_coefficients}, the propagation direction of the spin wave alters with the frequency $\omega_{\downarrow\uparrow}\equiv\omega_{\mathcal{C},q/2}-\omega_{\mathcal{S},q/2}\approx(G_2/2G_1)\sqrt{JG_1}(q/2)=(G_2/2G_1)\Omega$. For illustration, we plot in Fig. \ref{fig:static_G_2} the ``position of the wave" as a function of time at different $G_2$ by tracking the profile (e.g., the zero crossing) of the propagating wave.
We observe a number of zigzag curves describing the to-and-fro motion with slopes corresponding to the velocity of free spin waves. The propagation direction alters with a frequency, which is proportional to the magnitude of the deformation of the lattice induced by the strain. (From Fig. \ref{fig:static_G_2} we observe that at $G_1=2$ the picture works qualitatively well up to $G_2\lesssim0.9$ confirming the generality of the explanation.)

\begin{figure}[h]
\includegraphics[clip, width=1 \columnwidth]{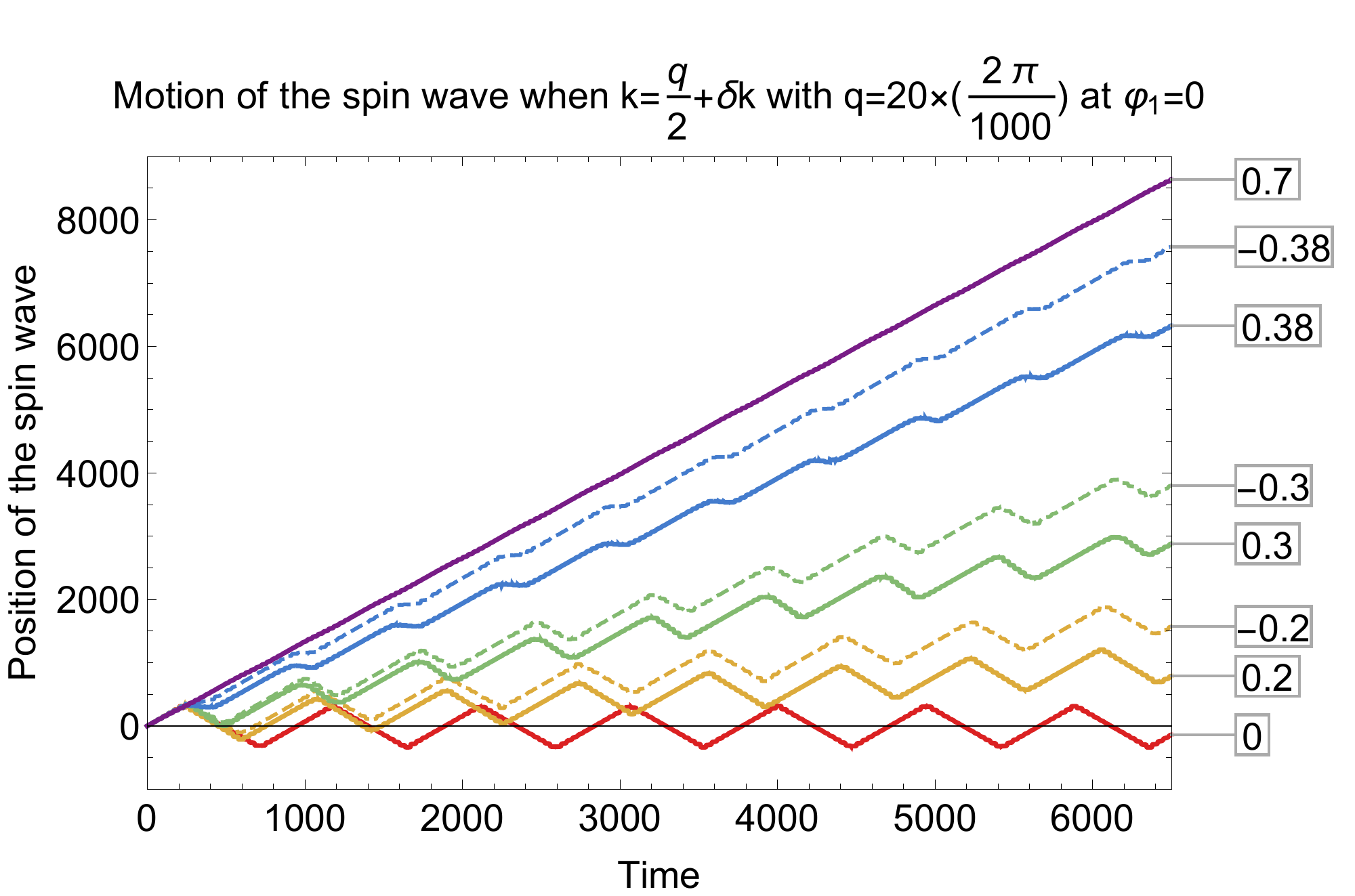}\llap{\makebox[7.55cm][l]{\raisebox{2.55cm}{\includegraphics[clip, width=0.33\columnwidth]{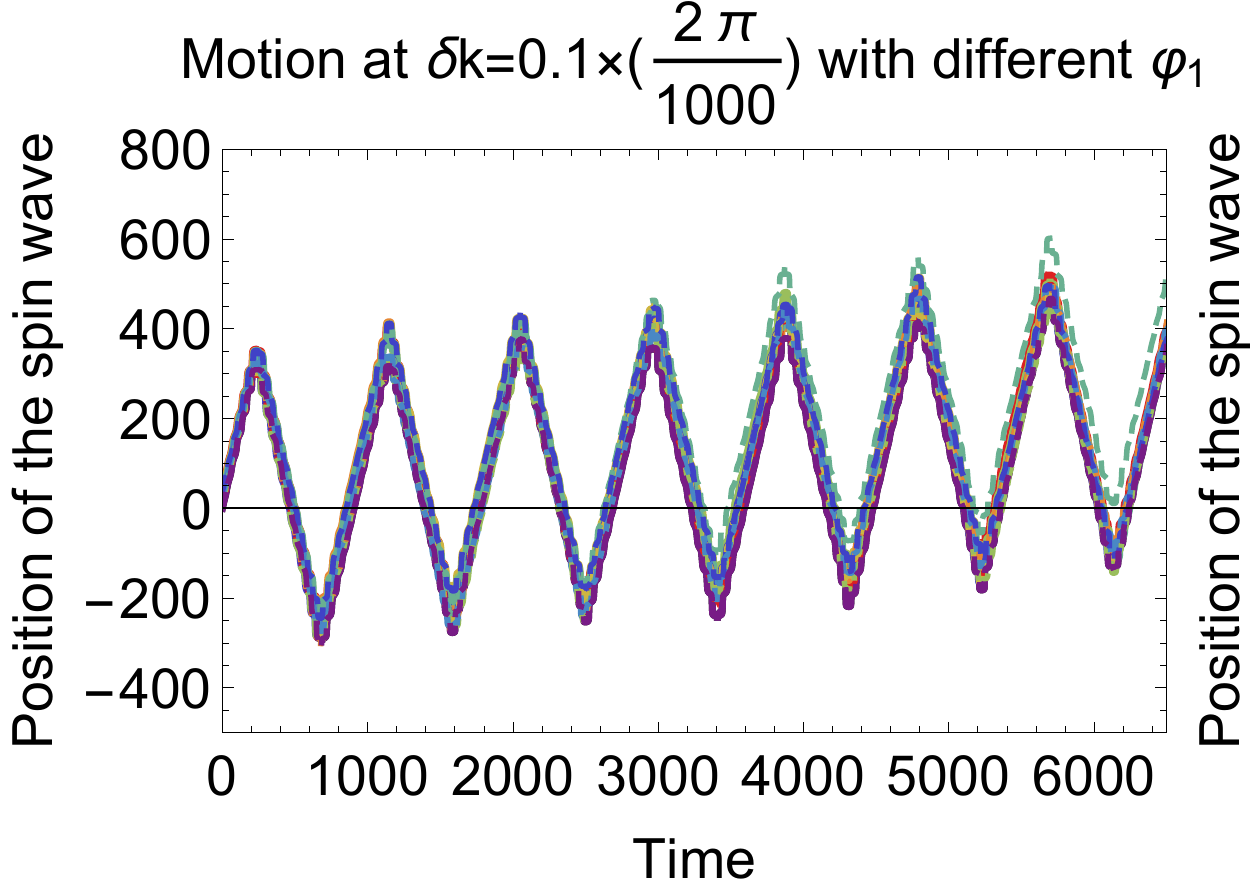}}}}
	
	\protect\caption{Propagation of spin waves with various $\delta k$ given in the unit $(2\pi/1000)$ (see boxes on the right). The parameters used for this simulation are $q=20\times(2\pi/1000)$, and $J=1$, $G_1=2$, $G_2=0.3$, $\varphi_1=0$. Inset is plotted for the wave at $\delta k=0.1\times(2\pi/1000)$ with different phases $\varphi_1$, which range from $-\pi$ to $\pi$.}
	
	\label{fig:static_delta_k}
\end{figure}

\emph{Out-of-resonance motion of the spin wave.}--When it comes to a slightly-out-of-resonance situation, i.e., $k=q/2+\delta k$ with $\delta k\neq0$ and $|\delta k|\ll q/2$, the approximated solution $\phi (z,t)$ contains 8 time-dependent components (which consist of two different quasimomenta $q/2\pm\delta k$, two opposite propagating directions, and two basis functions); see Sec. S2 of SM \cite{SM}. The results are presented in Fig. \ref{fig:static_delta_k}. First of all, one may notice that the durations of motion in the opposite directions are not equal anymore.
Furthermore, there is a critical value $\delta k_c$, so that for $|\delta k|>\delta k_c$, the to-and-fro motion of the spin wave ceases to exist. [For the discussed choice of the parameters, $\delta k_c\approx 0.38\times(2\pi/1000)$.]

\begin{figure}[ht] \centerline{\includegraphics[clip, width=1 \columnwidth]{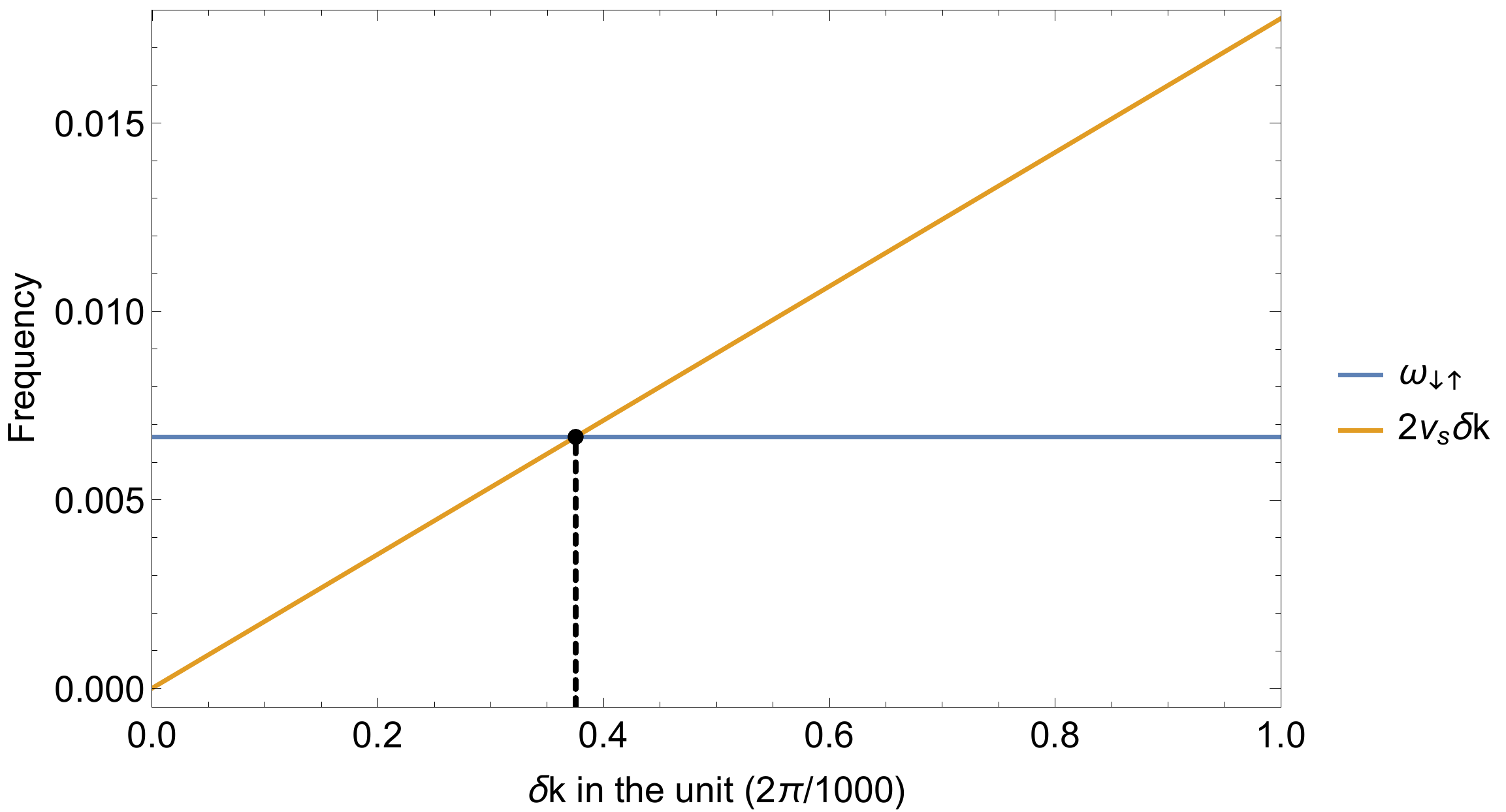}}

\protect\caption{The energy splitting of the free spin waves with $k=\frac{q}{2}\pm\delta k$, which are connected through the perturbation term $G_2\cos(qz)$. The orange line is the energy splitting, while the blue line is $\omega_{\downarrow\uparrow}$.}

\label{fig:static_energy_splitting}
\end{figure}

The critical value $\delta k_c$ can be determined by comparing the energy difference of the two spin waves connected through the external perturbation term $G_2\cos(qz)$, with the splitting energy $\omega_{\downarrow\uparrow}$. At $\delta k=\delta k_c$ the energy difference $2v_s\delta k$ meets the splitting energy $\omega_{\downarrow\uparrow}$; see Fig. \ref{fig:static_energy_splitting} (cf. tunneling in the slightly asymmetric double-well potential).

Next, in the inset of Fig. \ref{fig:static_delta_k}, we study the effect of the phase $\varphi_1$ in the initial conditions, which appears to be negligible. This can be explained as follows: As one can notice from Eq. (\ref{phi_static_MT}), the right- and left-propagating wave components contain $\varphi_1$ only in the combination $\pm(\Omega t-\varphi_1)$. By shifting the time with $t_0=\varphi_1/\Omega$, the phase is transferred to the arguments in the time-dependent coefficients. As a result, from $\omega_{\downarrow\uparrow}t/2$ they change to $\omega_{\downarrow\uparrow}(t+t_0)/2$, and therefore the effect of the shift leads only to a change in the moment of the turn of the propagating wave; cf. Fig. \ref{fig:static_time_dependent_coefficients}. (The same argument works for $\delta k\neq0$ as well; see Eq. (S23) in SM \cite{SM}). Furthermore, because of the smallness of $\omega_{\downarrow\uparrow}/2\Omega$, the effect appears to be negligible. Hence $\varphi_1$ practically does not affect the propagation of the spin wave.

This observation is of upmost importance. The absence of sensitivity to $\varphi_1$ implies that meeting of a spin-wave packet with the induced strain modulation can be considered instantaneous. In other words, the deformation effectively switches on for the whole wave packet simultaneously.

\emph{Further discussion.}--Finally, to demonstrate the generality of the idea, we considered an oblique incidence when the initial spin wave has a finite momentum component perpendicular to the direction of strain modulation. We observed that in this case the modulated strain acts like a waveguide; see Sec. S5 of SM \cite{SM} for the details. Namely, the to-and-fro motion develops along the direction of modulation, while in the direction perpendicular to the strain modulation the wave propagates freely.

For the sake of completeness, we also investigated the dynamics of spin wave (see Sec. S7 of SM \cite{SM}) in the presence of a time-dependent strain modulation $\epsilon_{zz}=\epsilon_0\cos(qz)\cos(\omega_{ph} t+\varphi_2)$. This can be achieved by a standing acoustic wave $\bm{u}=A\sin(qz)\cos(\omega_{ph} t+\varphi_2)\bm{e}_{z}$; see, e.g., Ref. [\onlinecite{chowdhury2017nondegenerate}]. Under the spatial resonance condition, we have observed that the to-and-fro motion can develop but is limited to the frequencies $\omega_{ph}\lesssim\omega_{\downarrow\uparrow}$, and its dynamics strongly depend on the phase $\varphi_2$.

\emph{Conclusions.}--In this letter, we discussed propagation of spin waves across the layered antiferromagnetic material in the presence of a static spatially modulated strain. We have found an alternating to-and-fro motion of the spin wave when its momentum is about half of the wave vector of the strain modulation, i.e., $k\approx q/2$ (we call it the \emph{spatial resonance} condition). The frequency of this to-and-fro motion $\omega_{\downarrow\uparrow}$ is proportional to the amplitude of the deformation.

As a practical application, this phenomenon can be used for controlling the spin-wave packets. Suppose a packet of spin waves centered around the wave vector $q/2$ is traveling freely across the layered magnetic system. Then, at a certain moment, one activates the modulated strain along the transverse direction with the quasi-momentum $q$. (Alternatively, perhaps more realistically for experimental realization, the wave packet runs inside the magnet when the deformation is switched on.) As follows from the discussion of Fig. \ref{fig:static_delta_k}, the Fourier components in the packet, which are closest to $q/2$, perform the to-and-fro motion, while the components that are more distant from $q/2$ pass through the sample. In this regard, the spatial modulation can work as a spin-wave filter and a delay line element.

\begin{acknowledgments}
We gratefully acknowledge the discussions with Brad Ramshaw and Arkady Shekhter. We thank Artem Abanov, Olena Gomonay, and Anatoli Polkovnikov for providing valuable comments.
\end{acknowledgments}

\bibliographystyle{apsrev}
\bibliography{MyBIB}

\end{document}